\documentclass[%
 reprint,
superscriptaddress,
 amsmath,amssymb,
 aps,
prapplied,
doi=true,
floatfix,
]{revtex4-1}

\usepackage[colorlinks=true,linkcolor=blue,urlcolor=blue,citecolor=blue]{hyperref}

\usepackage{graphicx}
\usepackage{dcolumn}
\usepackage{bm,upgreek}

\usepackage{overpic}
\usepackage{comment}
\usepackage{xfrac}
\usepackage{xcolor}
\usepackage{dblfloatfix}
\usepackage[mathscr]{euscript}
\newcommand{\mathcolorbox}[2]{\colorbox{#1}{$\displaystyle #2$}}
\usepackage[export]{adjustbox}
\definecolor{lightgray}{gray}{0.85}

\begin{document}

\preprint{APS/123-QED}

\title{Simultaneous Broadband Vector Magnetometry Using Solid-State Spins}

\author{Jennifer M. Schloss}
\affiliation{Department of Physics, Massachusetts Institute of Technology, Cambridge, Massachusetts 02139, USA}
\affiliation{Center for Brain Science, Harvard University, Cambridge, Massachusetts 02138, USA}
\author{John F. Barry}
\affiliation{Center for Brain Science, Harvard University, Cambridge, Massachusetts 02138, USA}
\affiliation{Lincoln Laboratory, Massachusetts Institute of Technology, Lexington, Massachusetts 02420, USA}
\affiliation{Harvard-Smithsonian Center for Astrophysics, Cambridge, Massachusetts 02138, USA}
\affiliation{Department of Physics, Harvard University, Cambridge, Massachusetts 02138, USA}
\author{Matthew J. Turner}
\affiliation{Center for Brain Science, Harvard University, Cambridge, Massachusetts 02138, USA}
\affiliation{Department of Physics, Harvard University, Cambridge, Massachusetts 02138, USA}
\author{Ronald L. Walsworth}
\email[Corresponding author.\\]{rwalsworth@cfa.harvard.edu}
\affiliation{Center for Brain Science, Harvard University, Cambridge, Massachusetts 02138, USA}
\affiliation{Harvard-Smithsonian Center for Astrophysics, Cambridge, Massachusetts 02138, USA}
\affiliation{Department of Physics, Harvard University, Cambridge, Massachusetts 02138, USA}

\date{\today}

\begin{abstract}

We demonstrate a vector magnetometer that simultaneously measures all Cartesian components of a dynamic magnetic field using an ensemble of nitrogen-vacancy (NV) centers in a single-crystal diamond. Optical NV-diamond measurements provide high-sensitivity, broadband magnetometry under ambient or extreme physical conditions; and the fixed crystallographic axes inherent to this solid-state system enable vector sensing free from heading errors. In the present device, multi-channel lock-in detection extracts the magnetic-field-dependent spin resonance shifts of NVs oriented along all four tetrahedral diamond axes from the optical signal measured on a single detector. The sensor operates from near DC up to a $12.5$\,kHz measurement bandwidth; and simultaneously achieves $\sim\!50$\,pT/$\sqrt{\text{Hz}}$ magnetic field sensitivity for each Cartesian component, which is to date the highest demonstrated sensitivity of a full vector magnetometer employing solid-state spins. Compared to optimized devices interrogating the four NV orientations sequentially, the simultaneous vector magnetometer enables a $4\times$ measurement speedup. This technique can be extended to pulsed-type sensing protocols and parallel wide-field magnetic imaging.

\end{abstract}

\pacs{Valid PACS appear here}
\maketitle

\section{\label{sec:intro}Introduction}
A wide range of magnetometry applications require real-time sensing of a dynamic vector magnetic field, including magnetic navigation~\cite{Webb1951,Lenz1990, Lenz2006, Cochrane2016}, magnetic anomaly detection~\cite{Lenz1990, Lenz2006}, surveying~\cite{GermainJones1957}, current and position sensing~\cite{Lenz1990,Lenz2006,Grosz2016}, and biomagnetic field detection and imaging~\cite{Wikswo1980, Hamalainen1993, Sander2012, LeSage2013, Glenn2015, Barry2016, Wang2017,Davis2018}. Scalar magnetometers, such as vapor cell, proton precession, and Overhauser effect magnetometers, measure only the magnetic field magnitude~\cite{Grosz2016}. Vector projection magnetometers, including SQUIDs, fluxgates, and Hall probes, measure the magnetic field projection along a specified axis in space; the determination of all three Cartesian field components then requires multiple sensors aligned along different axes. Uncertainty or drifts in the relative orientations or gains of these multiple sensors can result in heading errors, which limit the vector field reconstruction accuracy~\cite{Camps2009, Liu2014}. In contrast, the fixed crystallographic axes inherent to solid-state spin-based sensors allow complete vector field sensing while mitigating systematic errors from sensor axis misalignment and drifting gains~\cite{Maertz2010,Pham2011,LeSage2013,Lee2015,Niethammer2016, Cochrane2016}.

\begin{figure}[ht] 
\centering
\begin{overpic}[height=3.24 in]{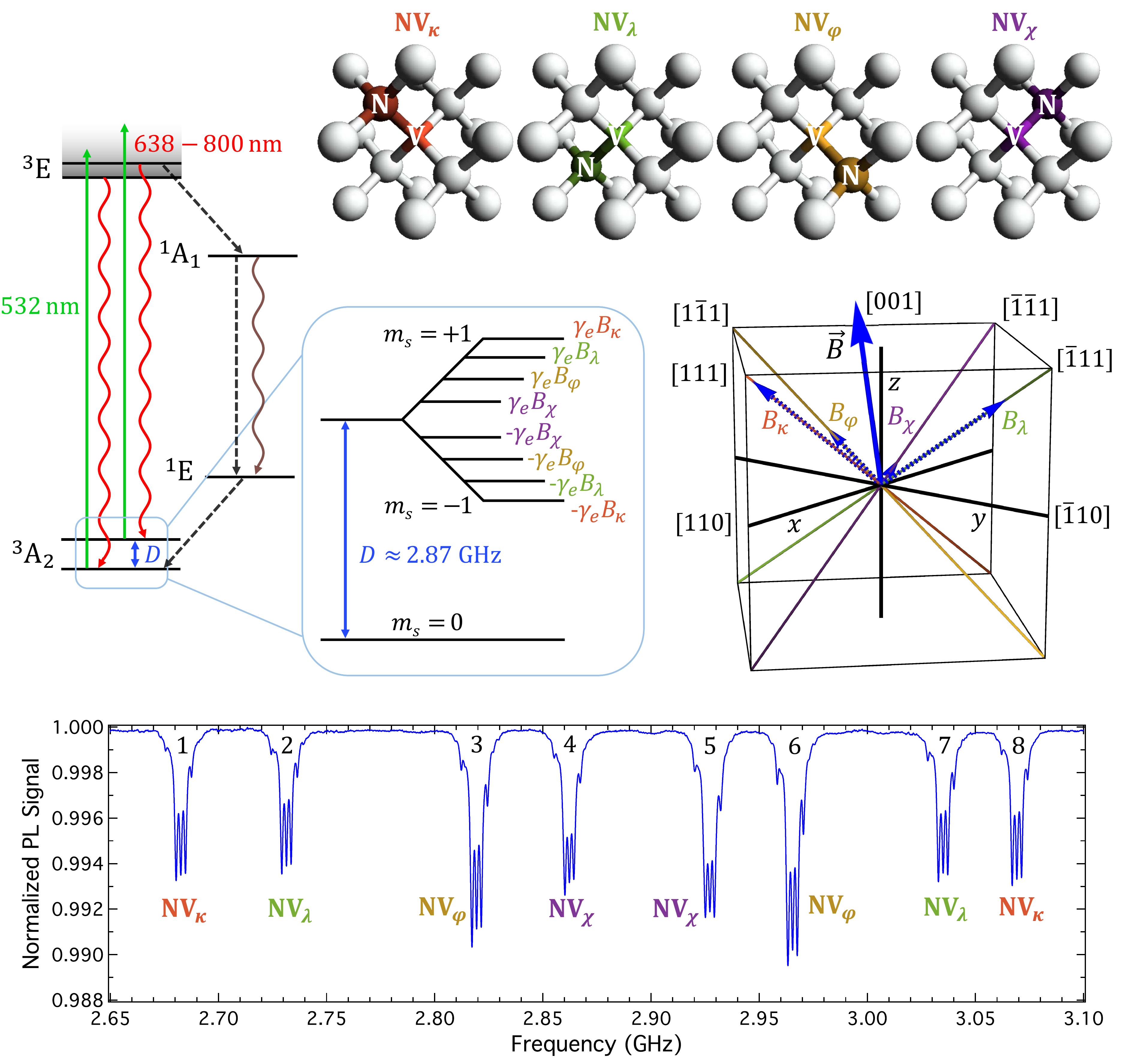}
\put (3.5, 87) {\small (a)}
\put (23, 88.75) {\small (b)}
\put (60.75, 69.5) {\small (c)}
\put (3.5, 34) {\small (d)}
\end{overpic}
\caption{(a) Energy level diagram for the nitrogen-vacancy (NV) center in diamond, with zero-field splitting $D$ between the ground-state electronic spin levels $m_s \!=\!0$ and $m_s \!=\!\pm 1$. Expanded view shows Zeeman shifts of the $m_s \!=\!\pm 1$ energy levels in the presence of a magnetic field $\vec{B}$ for different projections along the NV symmetry axis. (b) Four crystallographic orientations of the NV center in diamond. For an ensemble of NV centers within a single crystal diamond, the NV symmetry axes are equally distributed along the four orientations. (c) NV symmetry axes and lab-frame directions ($\hat{x}$, $\hat{y}$, $\hat{z}$), defined in terms of diamond lattice vectors. A magnetic field $\vec{B}$ projects onto the four NV orientations, causing the Zeeman shifts shown in (a). Shifts associated with off-axis magnetic fields are ignored for simplicity. (d) Optically detected magnetic resonance (ODMR) spectrum displaying photoluminescence (PL) signal from an ensemble of NV centers in a bias magnetic field $\vec{B} \!=\! \vec{B}_0 \!=\! \left(3.54, 1.73, 6.95\right)$\,mT, $|\vec{B}_0| \!=\! 7.99$\,mT. Resonance features numbered 1\,-\,4 (5\,-\,8) correspond to $|m_s\!=\!0\rangle \rightarrow |m_s\!=\!-1\rangle$ ($|m_s\!=\!0\rangle \rightarrow |m_s\!=\!+1\rangle$) spin transitions, and subfeatures arise from NV hyperfine structure (see Supplemental Material and Fig.~S3
)~\cite{Doherty2012, Felton2009}.}
\label{fig:intro}
\end{figure}
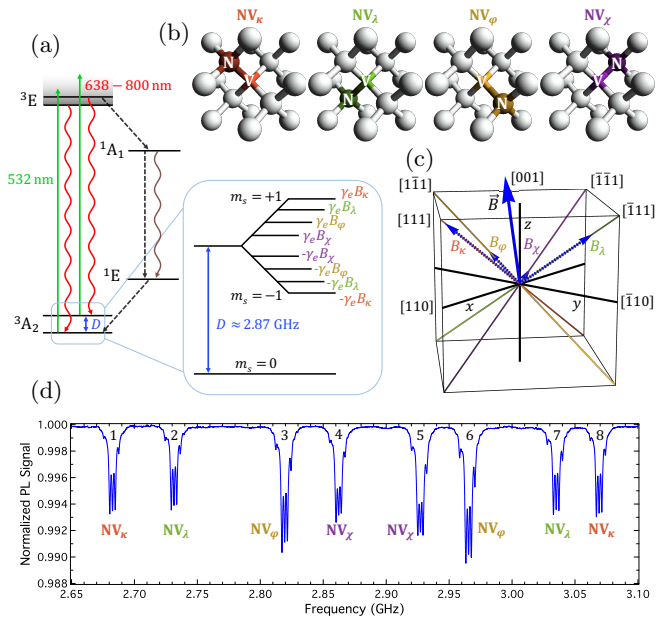

In particular, negatively charged nitrogen-vacancy (NV) centers in single-crystal diamond provide high-sensitivity broadband magnetic sensing and imaging under ambient conditions~\cite{Taylor2008}.  
The NV center is an atomic-scale defect consisting of a substitutional nitrogen adjacent to a vacancy in the lattice. The NV center's electronic ground state has spin $S\!=\!1$ with the lower-energy $m_s\!=\!0$ level separated from the $m_s\!=\!\pm 1$ levels by a zero-field splitting $D\!\approx \! 2.87$\,GHz (see Fig.~\ref{fig:intro}(a)). NV centers have symmetry axes aligned along one of four crystallographic orientations set by the tetrahedral symmetry of the diamond lattice (see Figs.~\ref{fig:intro}(b) and \ref{fig:intro}(c)). In a bias magnetic field $\vec{B}_0$, the $m_s\!=\!\pm 1$ energy levels of an NV center are Zeeman-shifted by $\approx \pm \hbar \gamma_e  \vec{B}_0 \cdot \hat{n}$ for fields $\gamma_e B_0\!\ll\!2 \pi D$, where $\gamma_e \!= \!g_e \mu_B/\hbar \!= \!2 \pi \times 28.03$\,GHz/T is the NV electron gyromagnetic ratio and $\hat{n}$ is the NV symmetry axis. The NV spin state can be prepared and read out optically, mediated by an intersystem crossing through a set of singlet states with preferential decay to the $|m_s\!=\!0\rangle$ state, causing higher photoluminescence (PL) from the $|m_s\!=\!0\rangle$ than from the $|m_s\!=\!\pm 1\rangle$ states~\cite{Goldman2015,Goldman2015b}. By measuring the optically detected magnetic resonance (ODMR) features of an $\it{ensemble}$ of NV centers, with NV symmetry axes distributed along all four crystallographic orientations (Fig.~\ref{fig:intro}(d)), the three Cartesian components of a vector magnetic field signal can be sensed using a monolithic diamond crystal~\cite{Pham2011}.

To date, ensemble NV vector magnetometers measure the three Cartesian magnetic field components either by sweeping a microwave (MW) tone across the full ODMR spectrum~\cite{Maertz2010,Steinert2010,Nowodzinski2015} or by interrogating multiple ODMR features, either individually~\cite{Taylor2008,Pham2011,LeSage2013,Wang2015a,Dmitriev2016, Blakley2016} or in parallel~\cite{Kitazawa2017,Zhang2018,Clevenson2018}, with near-resonant MWs. Although at least three ODMR features must be interrogated to determine the magnetic field vector, four or more are often probed to mitigate systematic errors from strain, electric fields, or temperature variation (see Appendix ~\ref{reconstruction})~\cite{Doherty2012,Dolde2011,Acosta2010}. Regardless of implementation, these existing methods all reconstruct the three Cartesian magnetic field components from a series of field projection measurements along at least three predetermined axes.

In existing implementations, the projective field measurements are performed sequentially, and such magnetometers have so far only demonstrated sensing of static or slowly varying fields fields~\cite{Maertz2010, Steinert2010, Pham2011, McGuinness2011, LeSage2013,Zhang2018,Clevenson2018}. A sequential vector magnetometer inherently exhibits suboptimal sensitivity, however, as the sensor is temporarily blind to magnetic field components \textit{transverse} to the chosen axis during each projective measurement. In addition, any dead time associated with the vector field measurement, including time spent switching the MW frequency or driving far off resonance, reduces the measurement speed and bandwidth as well as the achievable sensitivity of a shot-noise-limited device. 

To overcome these drawbacks, we demonstrate simultaneous measurement of all Cartesian magnetic field components using parallel addressing and readout from all four NV orientations. By performing four projective field measurements simultaneously, our technique enables high-sensitivity, broadband vector magnetometry without the inefficiency inherent to sequential projective measurement techniques. This method can decrease the time required to reconstruct a magnetic field vector with a given signal-to-noise ratio (SNR) by at least 4$\times$ compared to optimized sequential addressing of the NV orientations, resulting in at least $\sqrt{4} \!=\! 2 \times$ higher sensitivity for shot-noise-limited magnetometers. 


\section{Magnetometry Method}\label{magmethod}

In many high-sensitivity measurements, technical noise such as $1/f$ noise is mitigated by moving the sensing bandwidth away from DC via up-modulation. One method, common in NV-diamond magnetometry experiments, applies frequency~\cite{Shin2012, Barry2016, ElElla2017, Clevenson2018} or phase modulation~\cite{Pham2011, BarGill2013, Pham2016, BauchHart2018}
to the MWs addressing a spin transition, which causes the magnetic field information to be encoded in a band around the modulation frequency. Here we demonstrate a multiplexed extension of this scheme, where information from multiple NV orientations is encoded in separate frequency bands and measured on a single optical detector. Lock-in demodulation and filtering then extracts the signal associated with each NV orientation, enabling concurrent measurement of all components of a dynamic magnetic field.

In this technique, four dedicated MW tones, each dithered at a unique modulation frequency, address a subset of four of the eight ODMR features shown in Fig.~\ref{fig:intro}(d). The implementation here uses modulated continuous-wave (CW) ODMR, where the MWs are frequency modulated and the PL from all NV orientations is detected continuously on a single optical detector. Multi-channel demodulation and filtering in software reveal the ODMR line center shifts in response to a change in the magnetic field vector. The Cartesian components of the dynamic magnetic field are reconstructed in real time utilizing an approximated linear transformation derived from the NV ground state spin Hamiltonian (see Appendix~\ref{Bsens}). This simultaneous magnetometry method generalizes to addressing any number of ODMR features.

\section{Experiment}\label{experiment}

\begin{figure}[t] 
\centering
\begin{overpic}[height=2.55 in]{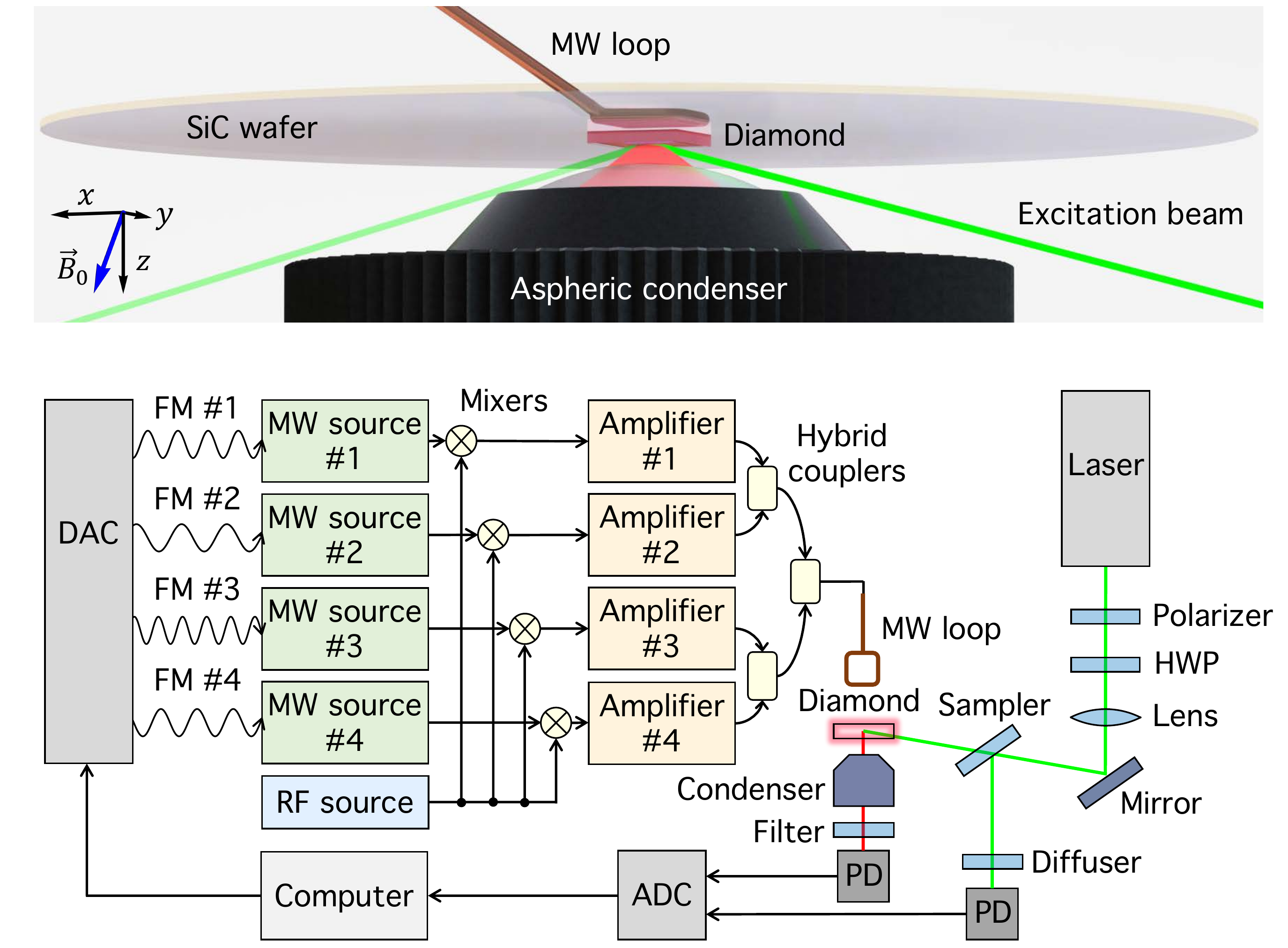}
\put (3.25, 70.4) {\small (a)}
\put (3.25, 44.6) {\small (b)}
\end{overpic}
\caption{a)~Laser and microwave (MW) excitation of NV centers in diamond sensor crystal and photoluminescence (PL) collection scheme. Diamond is affixed to one side of silicon carbide (SiC) wafer for stabilization and heat sinking. MW loop on reverse side of SiC provides modulated MW drive to NV ensemble. Excitation light at 532\,nm enters diamond at $\approx \! 73^\circ$ to the normal and PL is collected by aspheric aplanatic condenser shown below diamond. From this perspective, the $y$-axis points primarily into the page.
b)~Schematic of setup. Digital-to-analog converter (DAC) outputs MW frequency modulation (FM) waveforms. MWs are generated by four sources and mixed with radiofrequency (RF) signal at 2.158\,MHz to produce modulated carriers plus sidebands, which are amplified, combined, and radiated by the MW loop. Excitation laser beam passes through polarizer, half waveplate (HWP), and focusing lens. After a mirror, the beam passes through a beam sampler, where a fraction is imaged onto a photodiode and digitized at the analog-to-digital converter (ADC); and the rest of the beam impinges on the diamond. The diamond PL is collected by the aspheric condenser, long-pass filtered at 633\,nm, imaged onto a photodiode, and digitized. See Appendices~\ref{MWelectronics} and~\ref{optics} for additional information and Fig.~S1 of the Supplemental Material 
 for a detailed electronics schematic.} 
\label{fig:schematic}
\end{figure}

Figure~\ref{fig:schematic} depicts the experimental setup, including laser excitation, MW generation, and magnetic field detection. The diamond crystal is a 4\,mm$\,\times\,4\,$mm$\,\times\,0.5\,$mm chip with $\langle 110\rangle$ edges and a \{100\} front facet, with [$^{14}$N$]\approx 28$\,ppm and an N-to-NV$^-$ conversion efficiency of $\sim\!10\%$.
The diamond is adhered to a 2''\,diameter, $330\,\upmu$m thick semi-insulating silicon carbide (SiC) wafer for thermal and mechanical stabilization (see Fig.~\ref{fig:schematic}(a)). A 532\,nm, $\sim\! 3.3$\,W beam with 400\,$\upmu$m Gaussian $1/e^2$ width impinges on the diamond chip's \{100\} facet at $\approx  \! 73^\circ$ to the normal, exciting the NVs. An aspheric, aplanatic condenser collects the PL and directs it through a 633\,nm long-pass filter, after which $\sim\!52$\,mW of PL is imaged onto a photodiode and digitized, as shown in Fig.~\ref{fig:schematic}(b) (see Appendices~\ref{MWelectronics}-\ref{optics}). A picked-off portion of the green excitation light ($\sim \! 135$\,mW) is collected onto a second photodiode and digitized for software-based laser intensity noise cancellation (see Appendix~\ref{noisecancellation}).

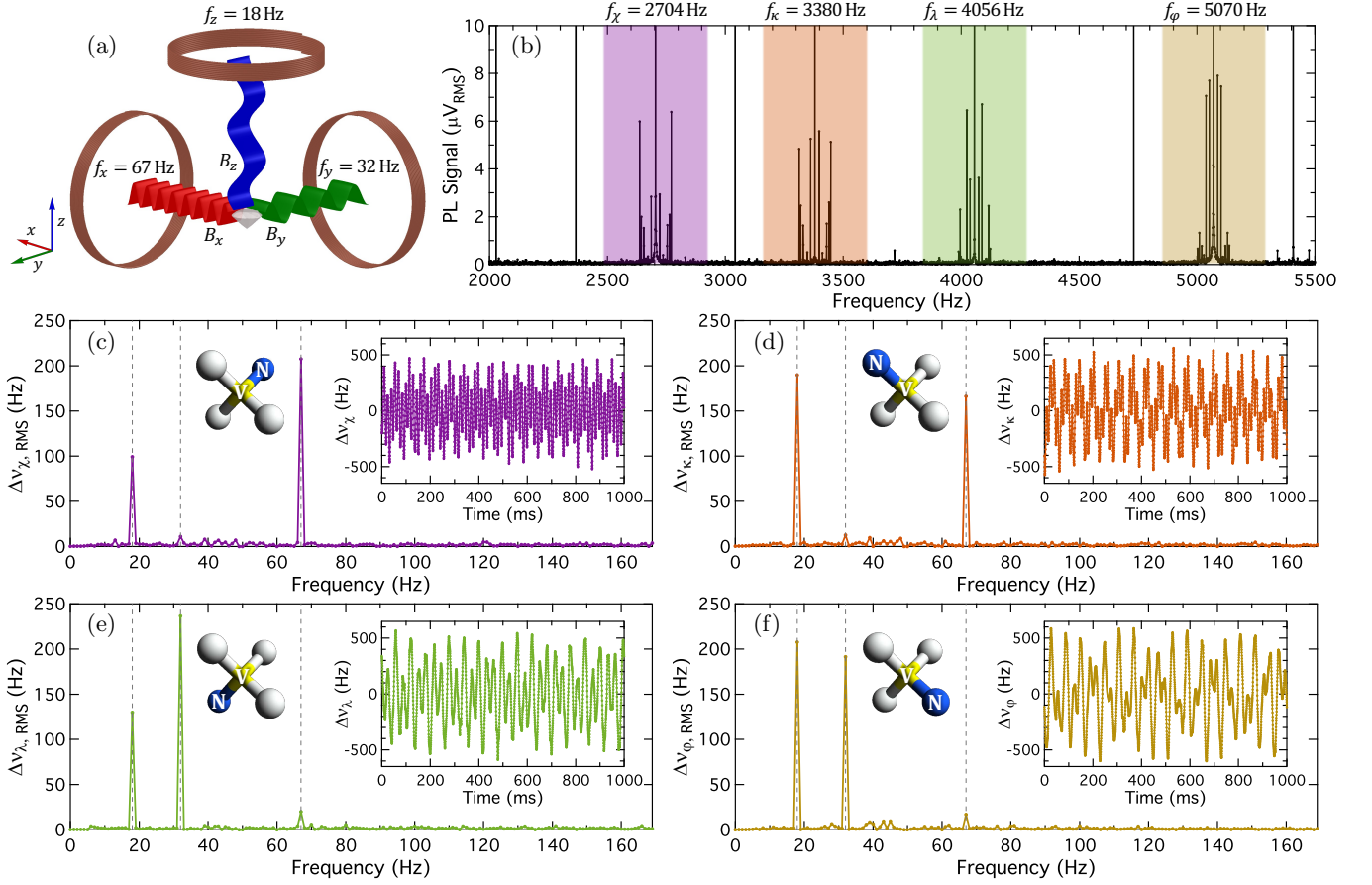
\begin{figure*}[t] 
\centering
\begin{overpic}[height=4.69 in]{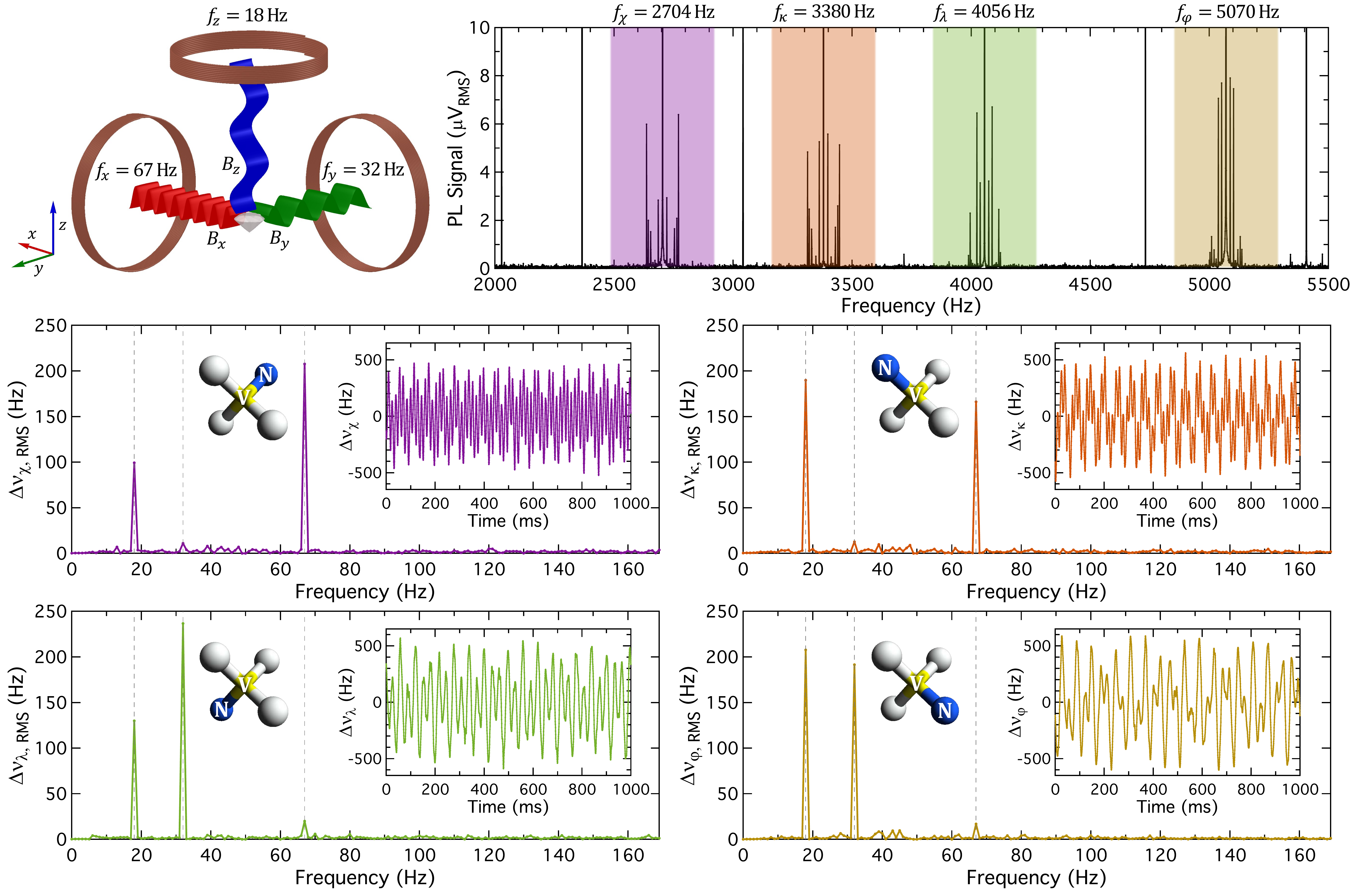}
\put (6.5,61.9) {\small (a)}
\put (38,61.9) {\small (b)}
\put (6.5,39.9) {\small (c)}
\put (56,39.9) {\small (d)}
\put (6.5,18.9) {\small (e)}
\put (56,18.9) {\small (f)}
\end{overpic}
\caption{Applied magnetic fields and simultaneous vector magnetometry data. a) Three coils generate magnetic field signals $B_x(t), B_y(t)$, and $B_z(t)$ at $f_x \!=\! 67$\,Hz, $f_y \!=\! 32$\,Hz, and $f_z \!=\! 18$\,Hz with root-mean-square (RMS) amplitudes $B_{x,\text{RMS}}\!=\!8.12$\,nT, $B_{y,\text{RMS}} \!=\! 9.56$\,nT, and $B_{z,\text{RMS}} \!=\! 9.86$\,nT. b) Spectral density of detected PL signal from 1 second of continuous acquisition. Shaded regions mark frequency bands containing four modulation frequencies $f_i$, where $i \!=\! \lambda, \chi, \varphi, \kappa$ (see Table~\ref{tab:params}), and encoded magnetic field signal around each $f_i$. Third-order intermodulation products at 2028\,Hz, 2366\,Hz, 3042\,Hz, and 4732\,Hz, and the second harmonic of $f_\chi$ at 5408\,Hz arise due to the nonlinear response of the NV PL to frequency-modulated MW driving (see Supplemental Material)~\cite{Lee2012}.
c-f) Spectral densities of magnetic-field-dependent frequency shifts $\Delta \nu_{i, \text{RMS}}$ of each addressed NV ODMR feature from 1 second of PL detection, after demodulating at $f_i$ and filtering; (inset), time series of same demodulated and filtered data, showing magnetic-field-dependent shifts $\Delta \nu_i$, and cartoons depicting the NV orientation corresponding to each measured $\Delta \nu_i$ trace.}
 \label{fig:nvshifts}
\end{figure*}

\begin{table*}[!b]
\begin{center}
    \begin{tabular}{ | c | c | c | c | c | c | c | } 
    \hline
    \textbf{MW Source} & \textbf{Carrier Freq.} & \textbf{NV Axis} & \textbf{Transition}  & \textbf{Mod.~Freq.} & \textbf{Freq.~Deviation} & \textbf{Lock-in Signal Slope} \\ \hline
 1 &   $\nu_\lambda = 2.731$\,GHz & $\hat{n}_\lambda \parallel [\bar{1}11] $ & $|0\rangle \leftrightarrow  |-$1$\rangle $ & $f_\lambda = 4056$\,Hz & $\delta\nu_\lambda = 832$\,kHz  &  $dS_\lambda/d\Delta \nu_\lambda = 39.5\,\upmu$V/kHz \\  \hline
  2 & $\nu_\chi = 2.862$\,GHz &  $\hat{n}_\chi \parallel [\bar{1}\bar{1}1]$ & $|0\rangle \leftrightarrow  |-$1$\rangle $ &  $f_\chi = 2704$\,Hz & $\delta\nu_\chi = 828$\,kHz  & $dS_\chi/d\Delta \nu_\chi = 42.0\,\upmu$V/kHz \\ \hline
      3 & $\nu_\varphi = 2.966$\,GHz & $\hat{n}_\varphi \parallel [1\bar{1}1]$ & $|0\rangle \leftrightarrow  |+$1$\rangle $ &  $f_\varphi = 5070$\,Hz & $\delta\nu_\varphi = 775$\,kHz  & $dS_\varphi/d\Delta \nu_\varphi = 53.4\,\upmu$V/kHz \\  \hline
          4 &   $\nu_\kappa = 3.069$\,GHz  & $\hat{n}_\kappa \parallel [111]$ & $|0\rangle \leftrightarrow  |+$1$\rangle $ & $f_\kappa = 3380$\,Hz  & $\delta\nu_\kappa = 1178$\,kHz &  $dS_\kappa/d\Delta \nu_\kappa = 41.6\,\upmu$V/kHz \\
    \hline
    \end{tabular}
    \caption{Parameters detailing the four-axis simultaneous vector magnetometry implementation, as discussed in the main text. \label{tab:params} }
\end{center}
\vspace{3 mm}
\hspace{75 mm}
\end{table*}

As shown in Fig.~\ref{fig:schematic}(b), four separate MW sources generate four carrier signals at frequencies $\nu_\lambda, \nu_\chi, \nu_\varphi$, and $\nu_\kappa$ resonant with the ODMR features respectively numbered 2, 4, 6, and 8 in Fig.~\ref{fig:intro}(d). These MW tones are sinusoidally frequency modulated at corresponding modulation frequencies  $f_\lambda, f_\chi, f_\varphi$, and $f_\kappa$ and frequency deviations $\delta\nu_\lambda, \delta\nu_\chi, \delta\nu_\varphi$, and $\delta\nu_\kappa$, all of which are tabulated in Table~\ref{tab:params}. A mixer for each carrier signal generates sidebands at $\pm 2.158$\,MHz to address hyperfine subfeatures~\cite{Barry2016}. These MW signals are amplified and combined onto a copper wire loop made from a shorted end of semi-rigid nonmagnetic coaxial cable, which is placed in proximity to the diamond, as depicted in Fig.~\ref{fig:schematic}(a). The modulation frequencies $f_i$, $i \!=\! \lambda, \chi, \varphi, \kappa$ are selected to balance measurement bandwidth and contrast~\cite{Shin2012,Barry2016} while ensuring that no frequency is an integer multiple of any other. The latter choice avoids cross-talk from the NVs responding nonlinearly to the modulated MWs at harmonics of $f_i$. All $f_i$ are chosen to divide evenly into the overall sampling rate $F_s \!=\! 202.800$\,kSa/s. Deviations $\delta \nu_\lambda$, $\delta \nu_\chi$, $\delta \nu_\varphi$, and $\delta \nu_\kappa$ are empirically optimized for maximal demodulated signal contrast. The optimal $\delta \nu_i$ depends on ODMR linewidth, which varies among the addressed NV resonances, as shown in Fig.~\ref{fig:intro}(d). The linewidth variations are the result of varying degrees of optical and MW power broadening, which arise from different laser and MW polarization angles with respect to the optical and magnetic transition dipole moments of the four NV orientations~\cite{Slichter1990, Herrmann2016, Backlund2017}. The demodulated signals $S_i$ are converted to ODMR line shifts $\Delta \nu_i$ via measured lock-in signal slopes $d S_i / d \Delta \nu_i $ reported in Table~\ref{tab:params}. These slopes are determined by sweeping the modulated MW carrier frequencies by 20\,kHz and fitting linear functions to the detected demodulated signals $S_i$ (see Appendix~\ref{slope}). 

As a demonstration, the NV-diamond sensor is placed in a bias magnetic field $|\vec{B}_0|\!=\! 7.99$\,mT, $\vec{B}_0 \!=\! \left(3.54, 1.73, 6.95\right)$\,mT, where the lab-frame coordinates $\left(x, y, z\right)$ are defined with respect to the normal faces of the mounted diamond crystal, with unit vectors $\hat{x}$, $\hat{y}$, and $\hat{z}$ lying along $[1 1 0]$, $[\bar{1} 1 0]$, and $[0 0 1]$, respectively, as depicted in Fig.~\ref{fig:intro}(c). In this coordinate system, the unit vectors parallel to the NV symmetry axes are $\hat{n}_\kappa \!=\! \left(\sqrt{\sfrac{2}{3}}, 0, \sqrt{\sfrac{1}{3}}\right) \!\parallel\! [111]$, $\hat{n}_\lambda \!=\! \left(0, -\sqrt{\sfrac{2}{3}}, -\sqrt{\sfrac{1}{3}}\right) \!\parallel \![\bar{1}11] $, $\hat{n}_\varphi \!=\! \left(0, \sqrt{\sfrac{2}{3}}, -\sqrt{\sfrac{1}{3}}\right) \!\parallel \![1\bar{1}1]$, and $\hat{n}_\chi \!=\! \left(-\sqrt{\sfrac{2}{3}}, 0, \sqrt{\sfrac{1}{3}}\right) \!\parallel\! [\bar{1}\bar{1}1]$. In this bias field $\vec{B}_0$, the ODMR spectrum is measured (see Fig.~\ref{fig:intro}(d)), and the resonance line centers are tabulated (see Appendix~\ref{bias}).  By numerically fitting the NV ground-state Hamiltonian to the eight measured resonance line centers (see Appendix~\ref{bias}), the bias field $\vec{B}_0$ is determined, along with the value of $D$ and the longitudinal electric/strain field coupling parameters $\vec{\mathscr{M}}_z\!\equiv\![\mathscr{M}_z^{\lambda},\mathscr{M}_z^{\chi}, \mathscr{M}_z^{\varphi}, \mathscr{M}_z^{\varphi}]$. From the fit, we obtain $D \!=\! 2.8692$\,GHz, $\vec{\mathscr{M}}_z\!=\!\left[-20, -60,~50,~30\right]$\,kHz, and $\vec{B}_0$ as reported above, which is consistent with a Hall probe measurement of $\vec{B}_0$ (see Appendix~\ref{appliedfields}).

Next, the Hamiltonian is linearized around the measured $\vec{B}_0$, $D$, and $\vec{\mathscr{M}}_z$, to calibrate the expected frequency shifts of the four addressed NV resonance line centers in the presence of an additional small field $\vec{B}_\text{sens}$ to be sensed. The result of the numerical linearization is a $4\times 3$ matrix $\textbf{A}$, where 
\begin{equation}\label{eqn:A}
\begin{bmatrix}
\Delta \nu_{\lambda}\\
\Delta \nu_{\chi}\\
\Delta \nu_{\varphi}\\
\Delta \nu_{\kappa}
\end{bmatrix}_\text{sens} = \textbf{A}
\begin{bmatrix}
B_{x}\\
B_{y}\\
B_{z}
\end{bmatrix}_\text{sens}.
\end{equation}
In the limit of small magnetic field and low strain, the rows of the matrix \textbf{A} are given, up to a sign, by the NV symmetry axis unit vectors $\hat{n}_i$ (see Appendix~\ref{Bsens}). The left Moore-Penrose pseudoinverse $\textbf{A}^{+}$ is then numerically calculated at the measured $\vec{B}_0$, $D$, and $\vec{\mathscr{M}}_z$ and used to transform detected frequency shifts $\Delta \nu_i(t)$, $i \!=\! \lambda, \chi, \varphi, \kappa$ (displayed in Figs.~\ref{fig:nvshifts}(c)-\ref{fig:nvshifts}(f)) into a measured vector field $\vec{B}_\text{sens} (t)$, shown in Fig.~\ref{fig:Bxyz}. This linearized matrix method provides a $\sim\!25,\!000\times$ speedup compared to unoptimized numerical least-squares fitting of the resonance frequencies $\nu_i$ in the presence of $\vec{B}_0 + \vec{B}_\text{sens}$. A comparison of the two methods yields good agreement, with fractional error $\lesssim 10^{-5}$ for sensed fields $\lesssim 100$\,nT (see Appendix~\ref{Bsens}).

\section{Vector Sensing Demonstration}\label{demo}

\begin{figure}[t] 
\centering
\begin{overpic}[height=4.52 in]{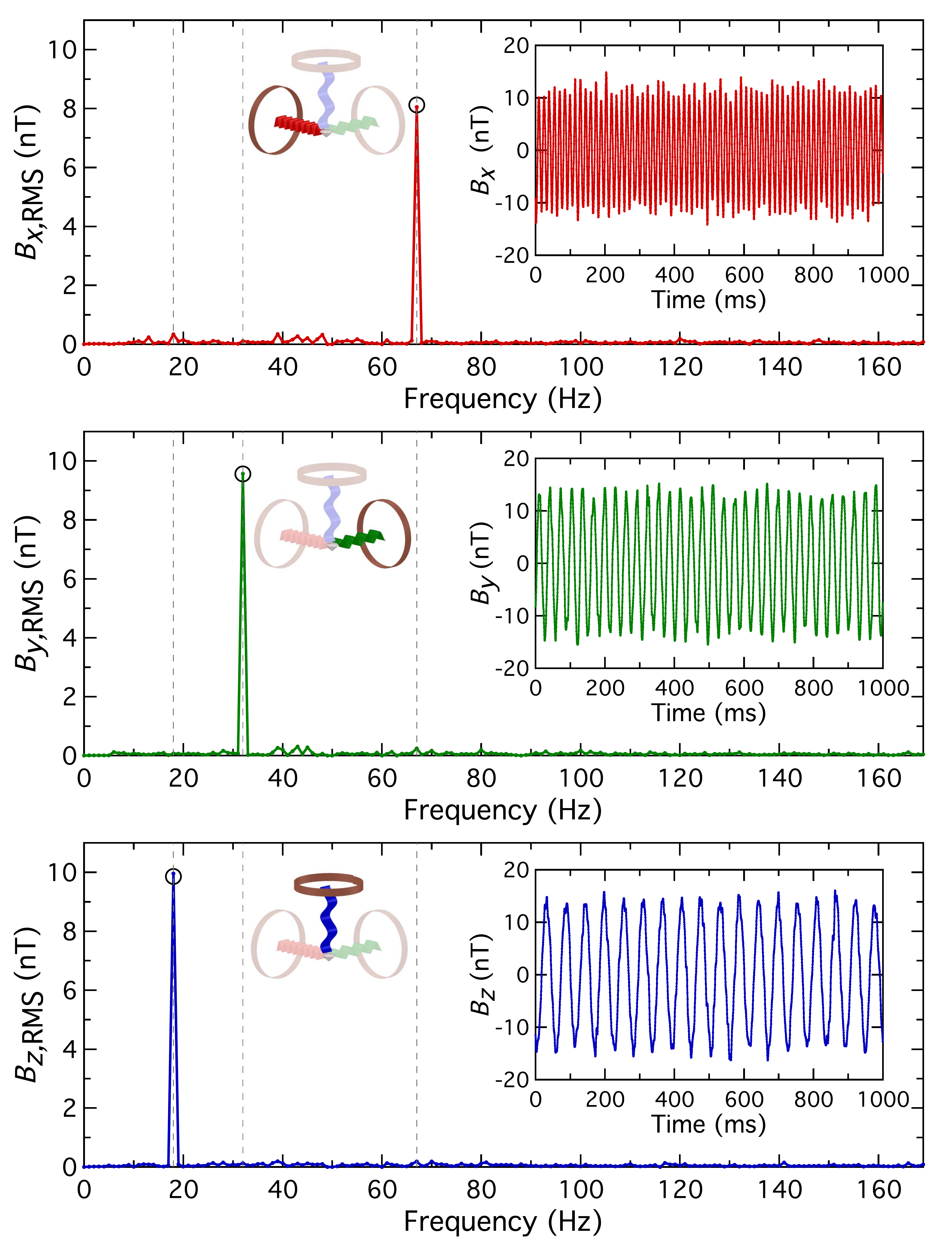}
\put (8.9,94.75) {\small (a)}
\put (8.9,61.75) {\small (b)}
\put (8.9,28.75) {\small (c)}
\end{overpic}
\caption{Detected magnetic fields using simultaneous vector magnetometer, extracted from data in Fig.~\ref{fig:nvshifts}. a) $B_x$ time trace (inset) and spectral density showing detected signal at $f_x \!=\! 67$\,Hz. b) $B_y$ time trace (inset) and spectral density showing detected signal at $f_y \!=\! 32$\,Hz. c) $B_z$ time trace (inset) and spectral density showing detected signal at $f_z \!=\! 18$\,Hz. Dashed lines mark applied signal frequencies $f_x$, $f_y$, $f_z$, and circles center on expected applied field amplitudes determined by sequential NV vector magnetometry. Cartoon recreations of Fig.~\ref{fig:nvshifts}(a) illustrate isolated detected components of dynamic vector magnetic field $B_x(t)$, $B_y(t)$, $B_z(t)$.} \label{fig:Bxyz}
\end{figure}

Three orthogonal coils create the time-varying magnetic field $\vec{B}_\text{sens}(t) \!=\! \left(B_x(t), B_y(t), B_z(t)\right)$ at the diamond sensor, where $B_j(t) \!=\!\sqrt{2} B_{j,\text{RMS}}\cdot \sin (2\pi f_j\cdot t + \phi_j)$ for $j \!=\! x, y, z$. Here $f_x \!=\! 67$\,Hz, $f_y \!=\! 32$\,Hz, $f_z \!=\! 18$\,Hz,  and the phases $\phi_x$, $\phi_y$, $\phi_z$ are chosen arbitrarily. The applied field amplitudes $B_{x,\text{RMS}} \!=\! 8.12$\,nT, $B_{y,\text{RMS}} \!=\! 9.56$\,nT, and $B_{z,\text{RMS}} \!=\! 9.86$\,nT are calibrated by conventional sequential NV magnetometry methods and are consistent with a priori calculations from the known coil geometries and applied currents. (See Supplemental Material for discussion of off-axis field nulling.) 

Figure~\ref{fig:nvshifts}(b) shows the voltage spectral density of the digitized, noise-canceled PL signal from 1 second of data acquisition. (See Fig.~S4 
 of the Supplemental Material for semi-log plots of the same data over different frequency ranges.)  The raw PL signal is high-pass filtered at 1690\,Hz and demodulated at the four modulation frequencies $f_i$, $i \!=\! \chi,\kappa,\lambda,\varphi$, by mixing the PL signal with a normalized sinusoidal waveform at each $f_i$. The four demodulated time traces are then band-pass filtered, notch-stop filtered, and downsampled to 2.704 kSa/s, producing the data shown in Fig.s~\ref{fig:nvshifts}(c)-\ref{fig:nvshifts}(f) (see Appendix~\ref{filtering}). The single-sided equivalent noise bandwidth of each of the resulting time traces is $f_\text{ENBW} \!=\! 203$\,Hz.  For applications requiring sensing at higher frequencies, measurement bandwidth can be greatly increased for a small (order unity) loss in sensitivity. (See Supplemental Material and Fig.~S2 
 for a demonstration with higher frequency magnetic fields and $\approx\!12.5$\,kHz measurement bandwidth.)

Figure~\ref{fig:Bxyz} displays the vector field components $B_x, B_y, B_z$ extracted from the measured frequency shifts of Figs.~\ref{fig:nvshifts}(c)-\ref{fig:nvshifts}(f). The extracted field components show good agreement with the amplitudes determined by sequential NV vector magnetometry, with differences at the $ 1 \%$ level or better (see Supplemental Material). Sensitivities $\eta_x$, $\eta_y$, and $\eta_z$ to magnetic field components along $\hat{x}$, $\hat{y}$, and $\hat{z}$ are determined from a series of magnetometry measurements with no applied magnetic signal. After multi-channel demodulation and filtering of these zero-signal traces, the detected frequency shifts $\Delta \nu_\lambda, \Delta \nu_\chi, \Delta \nu_\varphi, \Delta \nu_\kappa$ are extracted. These shifts are then transformed to $B_x, B_y$, and $B_z$ using the matrix $\textbf{A}^+$. The sensitivity $\eta_{j}$ to fields along the $j$ direction is given by 
\begin{equation}\label{eqn:sensitivity}
\eta_{j} = \frac{\sigma_{B_{j}}}{\sqrt{2f_\text{ENBW}}}
\end{equation}
for $j \!=\! x,y,z$, where $\sigma_{B_{j}}$ is the standard deviation of the zero-signal magnetic field time trace $B_{j}$. Sensitivities $\eta_x \!=\! 57$\,pT$/\sqrt{\text{Hz}}$,  $\eta_y \!=\! 46$\,pT$/\sqrt{\text{Hz}}$, and  $\eta_z \!=\! 45$\,pT$/\sqrt{\text{Hz}}$ are determined based on 1 second of recorded PL with $f_\text{ENBW} \!=\! 203$\,Hz. Photon shot-noise-limited sensitivities are calculated to be $\eta_x^\text{shot} \!=\! 18.1$\,pT/$\sqrt{\text{Hz}}$, $\eta_y^\text{shot} \!=\! 18.4$\,pT/$\sqrt{\text{Hz}}$, and $\eta_z^\text{shot} \!=\! 17.5$\,pT/$\sqrt{\text{Hz}}$ (see Appendix~\ref{shot}). The 2.5\,-\,3$\times$ factor above shot noise is attributed to uncanceled MW and laser intensity noise. The reported sensitivities are the highest demonstrated to date for any solid-state spin-based magnetometer performing broadband sensing of all magnetic field vector components.

\section{Proposed Pulsed Extension}\label{extensions}

\begin{figure}[t]
\centering
\begin{overpic}[height=3.5 in]{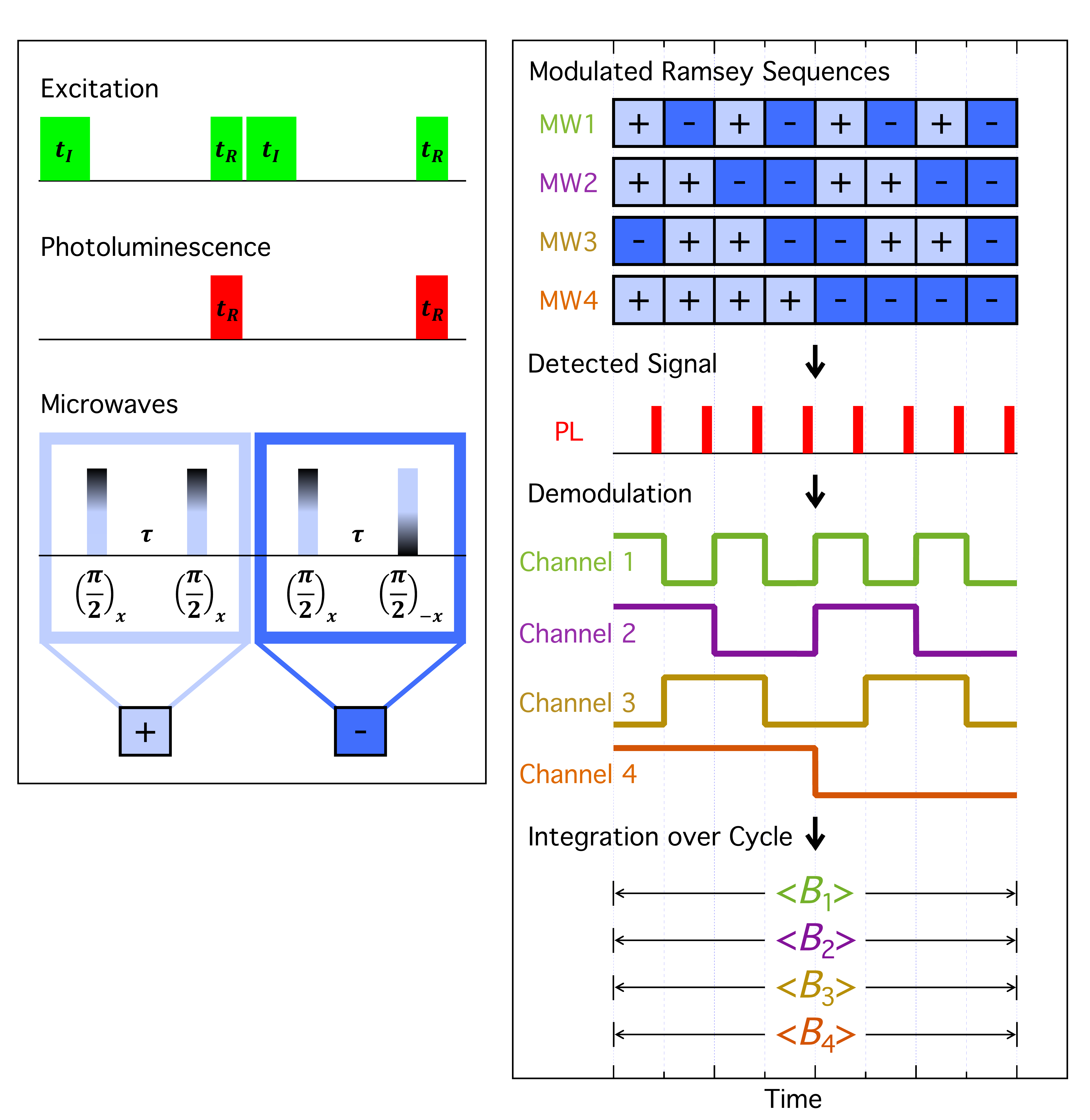} 
\put (2.5,98.2) {\small (a)}
\put (47,98.2) {\small (b)}
\end{overpic}
\caption{Proposed pulsed implementation of simultaneous vector magnetometry. a) Set of two Ramsey sequences with modulated MW phase. Green and red rectangles depict periods of 532\,nm laser excitation and PL collection, respectively, with NV-spin-state initialization time $t_I$ and spin readout time $t_R$. After an initial $\pi/2$ pulse, the magnetic field is sensed for a duration $\tau$  followed by a final MW $\pi/2$ pulse of variable phase. Sequences denoted $+$ and $-$ differ in phase by $180^\circ$ and yield equal and opposite PL contrast signals from a DC magnetic field~\cite{Pham2011,BarGill2013,DeVience2015}. 
b) Four-channel modulated Ramsey scheme. Final MW $\pi/2$ pulse phases from each sequence are modulated according to the a set of orthogonal Walsh codes~\cite{Walsh1923}. Detected PL signal is demodulated according to same Walsh codes to separate and extract magnetic-field-dependent shifts of four addressed NV orientations.}
\label{fig:pulsed}
\end{figure}

This simultaneous vector magnetometry method should be extendable to pulsed-type measurement protocols, such as Ramsey~\cite{Popa2004}, pulsed ODMR~\cite{Dreau2011}, and Hahn echo~\cite{vanOort1988}. For example, Ramsey magnetometry typically employs MW phase modulation in a modified variant of the dual measurement scheme used in NV sensing protocols to mitigate systematic noise sources~\cite{Pham2011,BarGill2013,DeVience2015}. In this dual measurement scheme, otherwise identical pulse sequences alternately project the final spin state onto the $|m_s\!=\!\pm 1\rangle$ and $|m_s\!=\! 0\rangle$ basis states by varying the phase of the final $\pi$/2 pulse, as illustrated in Fig.~\ref{fig:pulsed}(a). The magnetic field is then calculated using
\begin{equation}\label{eqn:Bpulsed2}
\langle B \rangle = \frac{\alpha}{2\langle S \rangle} \left[\mathcolorbox{white}{S_1} - \mathcolorbox{lightgray}{S_2}\right],
\end{equation}
where $S_i$ denotes the integrated PL signal resulting from the $i^\text{th}$ measurement, $\alpha$ is a proportionality constant between the magnetic field and the integrated PL, $\langle S \rangle$ denotes the mean integrated PL averaged over projections on the $|m_s\!=\!\pm 1\rangle$ and  $|m_s\!=\!0\rangle$ basis states, and the gray box (no box) indicates that the given measurement $S_i$ is projected onto the $|m_s\!=\!\pm 1\rangle$  ($|m_s\!=\!0\rangle$) state. The dual measurement scheme effectively removes background PL offsets, mitigates laser intensity fluctuations, and protects against certain systematics causing long-term drifts of $S_i$.

For the proposed extension to pulsed vector magnetometry, simultaneous pulse sequences are applied to multiple spectrally-separated ODMR features, each with a separate near-resonant MW frequency and a distinct alternation pattern of final $\pi$/2-pulse phases. Use of orthogonal binary sequences such as Walsh codes~\cite{Walsh1923,Gold1967,Beauchamp1984, Lathi1998, Cooper2013} for the phase alternation patterns ensures the detected PL can be demodulated to separate out the magnetic field signal associated with each NV orientation. For example, with the encoding scheme illustrated in Fig.~\ref{fig:pulsed}(b), these magnetic signals are given by
\begin{align}
\langle B_1 \rangle\! &= \!\frac{\alpha_1}{8 \langle S \rangle} \!\left[\;\;\;\!\mathcolorbox{white}{S_1}\!\!-\!\!\mathcolorbox{lightgray}{S_2}\!\!+\!\!\mathcolorbox{white}{S_3}\!\!-\!\!\mathcolorbox{lightgray}{S_4}\!\!+\!\!\mathcolorbox{white}{S_5}\!\!-\!\!\mathcolorbox{lightgray}{S_6}\!\!+\!\!\mathcolorbox{white}{S_7}\!\!-\!\!\mathcolorbox{lightgray}{S_8}\right]\label{eqn:B1}\\ 
\langle B_2 \rangle \!&= \!\frac{\alpha_2}{8 \langle S \rangle} \!\left[\;\;\;\!\mathcolorbox{white}{S_1}\!\!+\!\!\mathcolorbox{white}{S_2}\!\!-\!\!\mathcolorbox{lightgray}{S_3}\!\!-\!\!\mathcolorbox{lightgray}{S_4}\!\!+\!\!\mathcolorbox{white}{S_5}\!\!+\!\!\mathcolorbox{white}{S_6}\!\!-\!\!\mathcolorbox{lightgray}{S_7}\!\!-\!\!\mathcolorbox{lightgray}{S_8}\right]\label{eqn:B2}\\ 
\langle B_3 \rangle \!&= \!\frac{\alpha_3}{8 \langle S \rangle} \!\left[-\!\mathcolorbox{lightgray}{S_1}\!\!+\!\!\mathcolorbox{white}{S_2}\!\!+\!\!\mathcolorbox{white}{S_3}\!\!-\!\!\mathcolorbox{lightgray}{S_4}\!\!-\!\!\mathcolorbox{lightgray}{S_5}\!\!+\!\!\mathcolorbox{white}{S_6}\!\!+\!\!\mathcolorbox{white}{S_7}\!\!-\!\!\mathcolorbox{lightgray}{S_8}\right]\label{eqn:B3}\\ 
\langle B_4 \rangle \!&=\! \frac{\alpha_4}{8 \langle S \rangle} \!\left[\;\;\;\!\mathcolorbox{white}{S_1}\!\!+\!\!\mathcolorbox{white}{S_2}\!\!+\!\!\mathcolorbox{white}{S_3}\!\!+\!\!\mathcolorbox{white}{S_4}\!\!-\!\!\mathcolorbox{lightgray}{S_5}\!\!-\!\!\mathcolorbox{lightgray}{S_6}\!\!-\!\!\mathcolorbox{lightgray}{S_7}\!\!-\!\!\mathcolorbox{lightgray}{S_8}\right].\label{eqn:B4}
\end{align}

From the observed values of $\langle B_1 \rangle$, $\langle B_2 \rangle$, $\langle B_3 \rangle$, and $\langle B_4 \rangle$, the lab-frame magnetic field components $B_x$, $B_y$, and $B_z$ can be determined utilizing a linearized matrix as described previously (see Appendix~\ref{Bsens}). The simultaneous scheme (Eqns.~\ref{eqn:B1}-\ref{eqn:B4}) achieves the same bandwidth but a $2\times$ higher SNR than the scheme in Eqn.~\ref{eqn:Bpulsed2} applied sequentially to the four NV orientations, since both schemes require eight pulse sequences to reconstruct the magnetic field vector (see Supplemental Material). 

This pulsed implementation of simultaneous vector magnetometry is expected to allow improvements in both bandwidth and sensitivity compared to the demonstrated CW-ODMR implementation. In particular, sensing bandwidths up to $\sim \!100$\,kHz are anticipated, based on arguments in Ref.~\cite{Barry2016}. The two main contributors to the expected sensitivity enhancement are (i) more effective noise rejection due to modulated pulsed-type protocols encoding magnetic information at higher frequencies than CW-ODMR, and (ii) enhanced PL contrast from avoiding laser and MW power broadening~\cite{Dreau2011, Barry2016}.

\section{Outlook}\label{conclusion}

The method presented here allows simultaneous recording of all three Cartesian components of a dynamic vector magnetic field using a solid-state spin sensor. The technique is a straightforward extension of established methods for broadband magnetometry using ensembles of solid-state defects, and implementation in an existing system requires only additional MW components. The method offers at least a $2\times$ improvement in shot-noise-limited sensitivity, corresponding to a $4\times$ reduction in measurement time to achieve a target SNR when compared to sequential vector magnetic field sensing.  While the technique is demonstrated here for a single optical detector employing CW-ODMR, it is expected to be compatible with CW and pulsed-type measurements in both single-channel detectors and camera-based magnetic field imagers (see Appendix~\ref{camera}). 

\begin{acknowledgments}

This material is based upon work supported by, or in part by, the United States Army Research Laboratory and the United States Army Research Office under Grant No. W911NF1510548, as well as by the National Science Foundation Electronics, Photonics and Magnetic Devices (EPMD); Physics of Living Systems (PoLS); Integrated NSF Support Promoting Interdisciplinary Research and Education (INSPIRE) programs; Air Force Office of Scientific Research award FA9550-17-1-0371; and Lockheed Martin under award A32198. J.~M.~S.~was supported by a Fannie and John Hertz Foundation Graduate Fellowship and a National Science Foundation (NSF) Graduate Research Fellowship under Grant 1122374. We thank Pauli Kehayias, David Phillips, Mikael Backlund, Connor Hart, Erik Bauch, and David Glenn for helpful discussions. 

J.~M.~S.~and J.~F.~B.~contributed equally to this work, with J.~F.~B.~leading initial conceptualization and J.~M.~S.~leading execution of the experiments, analysis of the data, and paper writing. J.~M.~S., J.~F.~B., M.~J.~T., and R.~L.~W.~contributed to conception of the experiments.  J.~M.~S.~and J.~F.~B.~built the apparatus.  J.~M.~S.~conducted the experiments. J.~M.~S.~and M.~J.~T.~analyzed the results.  J.~M.~S., J.~F.~B., M.~J.~T., and R.~L.~W.~reviewed all results and wrote the paper.  R.~L.~W.~supervised the project.
\end{acknowledgments}

\appendix
\section{Imaging Implementation}\label{camera}

This simultaneous vector magnetometry method is expected to be compatible with camera-based wide-field magnetic imagers using NV-diamond~\cite{Steinert2010, Pham2011, LeSage2013, Chen2013, Steinert2013, Sarkar2014, Chipaux2015, Trusheim2016, Simpson2016, Glenn2017, Davis2018}, both for frequency-modulated CW-ODMR and for phase-modulated pulsed-type sensing schemes. We note that the demodulation and summation described in Section~\ref{extensions} and shown in Fig.~\ref{fig:pulsed} is a time-domain picture of the demodulation and low-pass filtering lock-in scheme used in the CW-ODMR demonstration. The same approach is expected to apply to camera-based parallel imaging, where the PL detection $S_n$ represents the $n^\text{th}$ camera exposure, and the signals $\langle B_1 \rangle,\langle B_2 \rangle,\langle B_3 \rangle,\langle B_4 \rangle$ and reconstructed field components $B_x,B_y,B_z$ represent magnetic field image frames. 

For frequency-modulated CW-ODMR magnetic imaging, square-wave modulation may enable increased SNR compared to sinusoidal modulation, as the adding and subtracting of image exposures amounts to square-wave demodulation of the detected signal. In both imaging and single-channel detection modalities, square-wave modulation and demodulation is expected to slightly increase measurement SNR by ensuring that the MWs always interrogate NV ODMR features at the points of steepest slope~\cite{Vanier1989,Barry2016,ElElla2017}.

\section{Vector Field Reconstruction}\label{reconstruction}

\subsection{Bias Field Determination}\label{bias}
The ODMR line center frequencies in the bias field $\vec{B}_0$ are measured by sweeping a single MW tone from 2.65 to 3.10\,GHz and recording the PL signal; this yields an ODMR spectrum as shown in Fig.~\ref{fig:intro}(d). Using a least-squares fit, we determine the line center of the middle hyperfine subfeature of each of the eight $m_s$ spin transitions. Averaging $10^3$ sweeps yields the following set of line centers, which are used to fit for the static field parameters $\vec{B} \!=\! \vec{B}_0$, $D$, and $\vec{\mathscr{M}}_z\!\equiv\![\mathscr{M}_z^{\lambda},\mathscr{M}_z^{\chi}, \mathscr{M}_z^{\varphi}, \mathscr{M}_z^{\varphi}]$:
\begin{equation}\label{eqn:linecenters}
\vec{\nu}_\text{ODMR} = \begin{bmatrix}
\nu_{\kappa^{-}}\\
\nu_{\lambda^{-}}\\
\nu_{\varphi^{-}}\\
\nu_{\chi^{-}}\\
\nu_{\chi^{+}}\\
\nu_{\varphi^{+}}\\
\nu_{\lambda^{+}}\\
\nu_{\kappa^{+}}
\end{bmatrix}=
\begin{bmatrix}
2.6825\\
2.7314\\
2.8193\\
2.8622\\
2.9272\\
2.9655\\
3.0351\\
3.0692
\end{bmatrix}\text{GHz}.
\end{equation}
A nonlinear least-squares (Levenberg-Marquardt) numerical minimization method is used to fit the differences between eigenvalues of the NV ground state spin Hamiltonian 
\begin{equation}\label{eqn:Hamiltonian}
\hat{\mathscr{H}}^{\,i} =h(D+\mathscr{M}_z^{\,i})(S_z^{\,i})^2+g_e \mu_B \vec{B}\cdot \vec{S}^{\,i}
\end{equation}
to the measured line centers for the four NV orientations $i \!=\! \lambda, \chi, \varphi, \kappa$, as described in the Supporting Information of Ref.~\cite{Glenn2017}. Here the dimensionless spin-1 operator of the NV triplet ground state $\vec{S}^{\,i}$ is defined in the NV body frame with $\hat{z} \equiv \hat{n}_i$; $D$ is the temperature- and strain-dependent zero-field splitting, defined specifically to be the coupling component $\propto (S_z^{\,i})^2$ that is common to all four NV orientations~\cite{Doherty2012}; and $\mathscr{M}_z^{\,i}$ is the additional anisotropic coupling, which differs between the four orientations and is attributed to longitudinal strain and electric fields~\cite{Barson2017, Glenn2017, Backlund2017}.
Coupling of transverse strain and electric fields $\mathscr{M}_x^{\,i}$ and $\mathscr{M}_y^{\,i}$ to the NV spin is suppressed by the on-axis component of the bias field, ($\frac{g_e \mu_B}{h}B_z^{\,i} \gg \mathscr{M}_x^{\,i}, \mathscr{M}_y^{\,i}$), and is therefore neglected~\cite{BauchHart2018,Jamonneau2016}. In contrast, the components of $\vec{B}_0$ transverse to the NV symmetry axes, ($B_x^{\,i}$ and $B_y^{\,i}$), contribute non-negligible shifts to the observed ODMR line centers. 

\subsection{Dynamic Magnetic Field Determination}\label{Bsens}

In the present demonstration, a subset of the detected ODMR line centers from Eqn.~\ref{eqn:linecenters} are selected for MW addressing $\vec{\nu}_\text{MW} \!=\! \vec{\nu}_\text{ODMR}\{2,4,6,8\} \!=\! [\nu_{\lambda^{-}},\nu_{\chi^{-}},\nu_{\varphi^{+}},\nu_{\kappa^{+}}]$. The $+$ and $-$ subscripts are dropped herein. The matrix $\textbf{A}$ from Eqn.~\ref{eqn:A}, reproduced here,
\begin{equation}\label{eqn:Areproduced}
\begin{bmatrix}
\Delta \nu_{\lambda}\\
\Delta \nu_{\chi}\\
\Delta \nu_{\varphi}\\
\Delta \nu_{\kappa}
\end{bmatrix}_\text{sens} = \textbf{A}
\begin{bmatrix}
B_{x}\\
B_{y}\\
B_{z}
\end{bmatrix}_\text{sens},
\end{equation}
is found by linearizing the Hamiltonian in Eqn.~\ref{eqn:Hamiltonian} about the measured $\vec{B}_0$, $D$, and $\vec{\mathscr{M}}_z\!\equiv\![\mathscr{M}_z^{\lambda},\mathscr{M}_z^{\chi}, \mathscr{M}_z^{\varphi}, \mathscr{M}_z^{\varphi}]$:  
\begin{equation}\label{Adef}
\textbf{A} =
\left.
\begin{bmatrix}
\frac{\partial \nu_\lambda}{\partial B_x} & \frac{\partial \nu_\lambda}{\partial B_y} & \frac{\partial \nu_\lambda}{\partial B_z}\\
\frac{\partial \nu_\chi}{\partial B_x} & \frac{\partial \nu_\chi}{\partial B_y} & \frac{\partial \nu_\chi}{\partial B_z}\\
\frac{\partial \nu_\varphi}{\partial B_x} & \frac{\partial \nu_\varphi}{\partial B_y} & \frac{\partial \nu_\varphi}{\partial B_z}\\
\frac{\partial \nu_\kappa}{\partial B_x} & \frac{\partial \nu_\kappa}{\partial B_y} & \frac{\partial \nu_\kappa}{\partial B_z}\\
\end{bmatrix}\right._{\vec{B}_0,D,\vec{\mathscr{M}}_z},
\end{equation}
where $B_x, B_y$, and $B_z$ are the lab-frame magnetic field components. The values of $D$ and $\vec{\mathscr{M}}_z$ are taken here to be constant during measurements, such that changes in the ODMR line centers are entirely attributed to magnetic field variations. 

The assumption of constant $\vec{\mathscr{M}}_z$ is valid in the present device because strain in the diamond is fixed and electric fields couple only very weakly to the NV energy levels at typical values of the bias magnetic field $\vec{B}_0$~\cite{vanOort1990,Jamonneau2016}. Although temperature drifts couple to $D$ with $dD/dT \!=\! 74$\,kHz/K~\cite{Acosta2010}, these drifts occur on timescales of seconds to hours, and the associated changes in $D$ are therefore outside the 5\,Hz to 210\,Hz measurement bandwidth of the present device. Furthermore, use of a SiC heat spreader attached to the diamond mitigates laser-induced temperature fluctuations (see Appendix~\ref{diamondmount})~\cite{Barry2016}. In a vector magnetometer optimized for sensing lower-frequency magnetic fields ($\lesssim$\,Hz), a changing zero-field splitting $D_\text{sens}$ could also be determined along with $\vec{B}_\text{sens}$ from the four measured ODMR line shifts. To additionally sense any dynamic changes in $\vec{\mathscr{M}}_z$ would require MW addressing of more than four ODMR features.

In the simple limiting case where, for each NV orientation $i \!=\! \lambda, \chi, \varphi, \kappa$, the transverse components of the bias magnetic field are much smaller than the zero-field splitting, ($B_x^{\,i},B_y^{\,i} \ll \frac{h}{g \mu_B}D$), and strain coupling is negligible, ($\mathscr{M}_x^{\,i}, \mathscr{M}_y^{\,i}, \mathscr{M}_z^{\,i}\ll \frac{g \mu_B}{h}B_z^{\,i}$), the marginal shifts $\Delta \nu_i$, are linearly proportional to the magnetic field projections $B_z^{\,i}$ along the respective NV symmetry axes. In this linear Zeeman regime, the rows of the matrix \textbf{A} are given, up to a sign, by the NV symmetry axis unit vectors $\hat{n}_i$:
\begin{equation}\label{eqn:Aproj}
\textbf{A}_\text{lin} = \frac{g_e \mu_B}{h}
\begin{bmatrix}
\hat{n}_\lambda\\
-\hat{n}_\chi\\
-\hat{n}_\varphi\\
\hat{n}_\kappa
\end{bmatrix}
=\frac{g_e \mu_B}{h}
\begin{bmatrix}
0 & -\sqrt{\sfrac{2}{3}} & -\sqrt{\sfrac{1}{3}}\\
\sqrt{\sfrac{2}{3}} & 0 & -\sqrt{\sfrac{1}{3}}\\
0 & -\sqrt{\sfrac{2}{3}} & \sqrt{\sfrac{1}{3}}\\
\sqrt{\sfrac{2}{3}} & 0 & \sqrt{\sfrac{1}{3}}
\end{bmatrix}.
\end{equation}
Whether the unit vector is multiplied by $+1$ or $-1$ depends on the sign of $B_z^{\,i} \equiv \vec{B}_0\cdot \hat{n}_i$ and on whether the addressed transition couples the $|m_s\!=\!\,0\!\,\rangle$ state to the $|m_s\!=\!\,+\!\,1\rangle$ or $|m_s\!=\!\,-\!\,1\rangle$ state.

Because the bias field $\vec{B}_0$ has non-negligible components transverse to each NV symmetry axis, the present experiment does not satisfy the requirement that only magnetic field projections on the NV symmetry axes contribute to the measured ODMR frequency shifts, and consequently \textbf{A} differs from $\textbf{A}_\text{lin}$. We determine \textbf{A} numerically by evaluating the partial derivatives using a step size of $\delta B_x \!=\! \delta B_y \!=\! \delta B_z \!=\! \frac{h}{g_e \mu_B}\cdot 10$\,Hz. The matrix \textbf{A} is calculated to be
\begin{equation}\label{eqn:Avalues}
\textbf{A} = \frac{g_e \mu_B}{h}
\begin{bmatrix}
0.10388 & -0.89383 & -0.46341 \\
0.90435  &  0.04596 & -0.38836\\
0.10524 &  -0.69511 &  0.73528\\
0.75551 &  0.04984 &   0.66268
\end{bmatrix}.
\end{equation}
The Moore-Penrose left pseudoinverse $\textbf{A}^{+}$ used in the experiment to determine $\vec{B}_\text{sens} (t)$ is numerically computed from $\textbf{A}$ with the MATLAB function \texttt{pinv} and is found to be  
\begin{equation}
\textbf{A}^{+} = \frac{h}{g_e \mu_B}
\begin{bmatrix}
0.08016 & 0.69252 & -0.02239 & 0.48676\\
   -0.71456  &  0.05848 &  -0.50880    & 0.09912\\
   -0.39850 &  -0.37710 &   0.51861   & 0.43394\\
\end{bmatrix}.\label{eqn:Aplus}
\end{equation}
The entries of the matrix $\textbf{A}^{+}$ are robust to small variation in the bias magnetic field and other Hamiltonian parameters. A 10\,$\upmu$T change in $B_x$, $B_y$, or $B_z$ changes no entry of $\textbf{A}^{+}$ by more than 1\% and some by less than 0.01\%. Doubling the strain parameters $\vec{\mathscr{M}}_z$ also affects the entries of $\textbf{A}^{+}$ by less (and for most entries much less) than 1\%. A 150\,kHz change in $D$ (corresponding to a temperature change of 2\,K~\cite{Acosta2010}) affects the entries of $\textbf{A}^{+}$ by 0.01\% or less. Thus, drifts in temperature or in the bias electric, strain, or magnetic fields have a negligible effect on the reconstruction accuracy of $\vec{B}_\text{sens}$ and can be ignored.

This linearized matrix method was compared against a numerical least-squares Hamiltonian fit, which is identical to the eight-frequency minimization method for $\nu_\text{ODMR}$ described in Appendix~\ref{bias} except that $D$ and $\vec{M}_z$ are held constant and $\vec{B} \!=\! \vec{B}_0+ \vec{B}_\text{sens}$ is determined from only the four frequencies $\nu_\text{MW}$. The methods were compared for a range of simulated fields $\vec{B}_\text{sens}$ ranging from 1\,nT to 100\,$\upmu$T. The linear transformation using $\textbf{A}^+$ from Eqn.~\ref{eqn:Aplus} agreed with the Hamiltonian fit to better than 0.001\% for $|\vec{B}_\text{sens}|\lesssim 100$\,nT and to better than 0.3\% for $|\vec{B}_\text{sens}|\lesssim 100$\,$\upmu$T.  When run on the same desktop computer, the linearized matrix method determines $\vec{B}_\text{sens}$ from $\vec{\nu}_\text{MW}$ with $\sim\!20\,\upmu$s per measurement, whereas the least-squares Hamiltonian fit 
requires $\sim\!500$\,ms per field measurement. This $\sim\!25,\!000\times$ speedup enables real-time vector magnetic field reconstruction from sensed frequency shifts in the present device.

\section{Experimental Setup}\label{setup}

\subsection{Microwave Electronics}\label{MWelectronics}
Figure~S1 
of the Supplemental Material shows the electronic equipment used in the present experiment. Agilent E8257D, E4421B, E4421B, and E4422B MW synthesizers generate the four modulated carrier frequencies $\nu_\lambda, \nu_\chi, \nu_\varphi$, and $\nu_\kappa$, respectively. These carriers are sinusoidally frequency modulated at $f_\lambda, f_\chi, f_\varphi$, and $f_\kappa$ with frequency deviations $\delta\nu_\lambda, \delta\nu_\chi, \delta\nu_\varphi$, and $\delta\nu_\kappa$ using the two analog outputs of each of two National Instruments (NI) PXI 4461 cards within an NI PXIe-1062Q chassis. The chassis also contains another NI PXI-4461 card used to trigger MW sweeps on the Agilent E8275D and for digitizing the detected optical signals, and an NI PXI-4462 card also for digitizing optical signals. All cards within the chassis and all MW and RF sources are synchronized to the Agilent E8257D's 10\,MHz clock using a distribution amplifier (Stanford Research Systems FS735). 

An Agilent E4430B synthesizer generates a 2.158\,MHz RF tone for the MW sidebands. A power divider (TRM DL402) splits this 2.158\,MHz signal four ways. Each of the split signals passes through a Mini-Circuits SLP-19+ low-pass filter and then is mixed with one of the four modulated carrier MW signals using a Relcom double-balanced mixer, either M1J or M1K. Before the mixer, each carrier signal passes through a Teledyne 2-4.5\,GHz isolator and a 10\,dB directional coupler. The coupled portion of the carrier signal passes through a 3\,dB attenuator and is then combined (Mini-Circuits ZX10-2-42-S+) with the sideband frequencies after the mixer to generate a MW signal resonant with all three hyperfine-split spin resonances for a given ODMR feature~\cite{Doherty2012, Barry2016}. 

The four sets of modulated MWs are amplified using four Mini-Circuits ZHL-16W-43-S+ amplifiers. The amplifier outputs pass through Teledyne 2-4.5\,GHz isolators and then circulators (either Pasternack PE 8401 or Narda 4923, each terminated with a 10\,W 50\,$\Omega$ terminator), before being combined with each other using three hybrid couplers (two Anaren 10016-3 and one Narda 4333, each terminated with a 10\,W 50\,$\Omega$ terminator). The combined MWs pass through a 20\,dB directional coupler, which picks off a portion of the signal for monitoring on a spectrum analyzer, and the rest is sent to a copper wire loop near the diamond to drive the NV spin resonances. The loop is positioned so that MWs are polarized approximately along $\hat{z}$ in the lab frame and address the four NV orientations roughly equally. 

\subsection{Diamond Mounting}\label{diamondmount}

The diamond is a 4\,mm\,$\times$\,4\,mm\,$\times$\,500\,$\upmu$m chip with $\langle 110\rangle$ edges and a \{100\} front facet, grown by Element Six Ltd. The diamond contains a bulk density of grown-in nitrogen $[^{14}\text{N}]\!\approx\!4.9\!\times\!10^{18}$\,cm$^{-3}$ and an estimated N-to-NV$^-$ conversion efficiency of $\sim\!10\%$ after irradiation and annealing. The diamond is affixed to a 2'' diameter, 330\,$\upmu$m thick wafer of semi-insulating 6H silicon carbide (SiC) from PAM-XIAMAN to provide both thermal and mechanical stability to the diamond crystal. The SiC wafer is in turn affixed to a 0.04'' thick tungsten sheet for additional mechanical stability, which is attached to an aluminum breadboard. The breadboard is mounted vertically so that the SiC and diamond surface (\{100\} face) make a $90^\circ$ angle with the optical table. A 1.5'' hole cut in the center of the tungsten sheet and a 2'' aperture in the center of the breadboard enable the SiC and diamond to be accessed from both sides. The axis normal to the diamond surface, aligned with the $[001]$ crystal lattice vector, is defined to be the lab-frame $z$-axis; the vertical axis normal to the optical table and along crystal lattice vector $[\bar{1} 1 0 ]$ is the lab-frame $y$-axis, and the horizontal axis perpendicular to both the $z$- and $y$-axes, along $[ 1 1 0 ]$, is the $x$-axis (see Fig.~\ref{fig:intro}(c)). 

\subsection{Applied Magnetic Fields}\label{appliedfields}
The bias magnetic field $\vec{B}_0$ is applied via a pair of 50\,mm diameter, 30\,mm thick N52 NdFeB neodymium magnets (Sunkee). The bias field is measured by sweeping a MW tone over the ODMR spectrum, as described in Appendix~\ref{bias}, and is consistent with a Hall probe measurement of $|\vec{B}_0|\!=\!8.4$\,mT, which is itself limited by the Hall probe's precision and $\sim$\,cm standoff distance from the diamond. 

The magnetic field coils are displaced from the diamond along the $x$-, $y$-, and $z$-axes by 28\,cm, $-$26.5\,cm, and 28.5\,cm respectively. The coils respectively contain 129, 130, and 142 turns and have diameters of 25\,cm, 32\,cm, and 25\,cm. Sinusoidally varying currents with root-mean-square (RMS) amplitudes of 0.24\,mA, 0.13\,mA, and 0.28\,mA  were applied to the respective coils. Based on the coil geometries and placement, and assuming that the coils are centered on the diamond and that there is no distortion of the fields by the magnetizable steel optical table, (an incorrect assumption, as discussed in the Supplemental Material), 
we expect $B_{x,\text{RMS}} \!=\! 10.51$\,nT, $B_{y,\text{RMS}} \!=\! 8.71$\,nT, and $B_{z,\text{RMS}} \!=\! 12.86$\,nT. The magnetic fields at the diamond were also determined via conventional sequential vector magnetometry, which found $B_{x,\text{RMS}} \!=\! 8.12$\,nT, $B_{y,\text{RMS}} \!=\! 9.56$\,nT, and $B_{z,\text{RMS}} \!=\! 9.86$\,nT, as marked by the open circles in Fig.~\ref{fig:Bxyz}. The discrepancies between the measured fields and the fields predicted from the known geometry and applied currents are attributed to magnetic field distortion by the ferromagnetic optical table, which is confirmed by a numerical simulation (Radia)~\cite{Elleaume1997,Chubar1998} (see Supplemental Material).

\subsection{Optical Setup}\label{optics}
The excitation laser source is a 532\,nm Verdi V-5 outputting 4.3\,W during typical operating conditions. The laser output passes through a Glan-Thompson polarizer, a half waveplate, and an $f\!=\!400$\,mm focusing lens. A silver mirror (Thorlabs PF10-03-P01) then directs the beam through a beam sampler (Thorlabs BSF10-A), after which 3.3\,W impinges on the \{100\} diamond surface at an oblique angle $\approx \!73^\circ$ to the normal. Reflections and scattered excitation light are reflected back toward the diamond using an aluminized mylar sheet opposite the excitation light entry side. The PL from the diamond is collected by an aspheric, aplanatic condenser (Olympus 204431), is long-pass filtered at 633\,nm (Semrock LP02-633RU-25), and $\sim\!52$\,mW is imaged onto a silicon photodiode (Thorlabs FDS1010), termed the \textit{signal photodiode}. This photodiode is reversed biased at 25 volts with a voltage regulator (Texas Instruments TPS7A49) followed by two capacitance multipliers in series~\cite{Hobbs2009}. The photocurrent is terminated into $R_\text{sig} \!=\! 300\,\Omega$. Before the diamond, the beam sampler picks off and directs $\sim\!135$\,mW of the excitation light through a beam diffuser and onto a second identical photodiode, termed the \textit{reference photodiode} (see Fig.~\ref{fig:schematic}(b)). This photodiode is powered from the same voltage source as the signal photodiode, and its photocurrent is terminated into $R_\text{ref} \!=\! 270\,\Omega$. 

Each photodiode voltage signal is simultaneously digitized by three analog-to-digital converters (ADCs), two of which are AC-coupled channels of an NI PXI-4462 digitizer. The signals from the AC-coupled channels are averaged in software to reduce digitization noise. Each photodiode voltage signal is also digitized by third ADC, which is a DC-coupled channel of an NI PXI-4461 digitizer. The signals from the DC-coupled channels are used to implement the laser noise cancellation. For the data shown in Figs.~3 and 4, all channels operate at sampling rate $F_s \!=\! 202.8$\,kSa/s.

\subsection{Laser Noise Cancellation}\label{noisecancellation}
Laser intensity noise is canceled by scaling and subtracting the green reference signal from the diamond PL signal. The digitized AC voltage from the reference photodiode, sampled over a 1-second-long interval, is scaled and subtracted from the AC voltage from the signal photodiode. The scaling factor for each interval is the ratio of the signal photodiode's mean DC value to the reference photodiode's mean DC value from that interval. This cancellation reduces the experimental noise in the 2-6\,kHz frequency band by $\sim\!30 \times$, achieving a noise level (in the absence of MW noise) that is $\sim\!1.5 \times$ above the expected level due to shot noise from both the signal and reference photocurrents (see Appendix~\ref{shot}). Under operating conditions with all modulated MWs in use, the experimental noise floor is 2.5\,-\,3$\times$ above this same expected shot-noise level.

\subsection{Demodulation and Filtering}\label{filtering}
As described in the main text, lock-in demodulation is performed in software in the present implementation, although the technique is also compatible with hardware demodulation. The noise-canceled PL signal is first high-pass filtered at 1690\,Hz with a 10th-order Butterworth filter, then separately mixed with sinusoidal waveforms at the four modulation frequencies given in Table~\ref{tab:params}.

The four demodulated traces are band-pass filtered (5\,Hz to 210\,Hz 10th-order Butterworth filter), which eliminates cross-talk from the other channels, upmodulated signal at $2f_i$, and environmental noise outside the sensing band. The filtered traces are then downsampled (decimated without averaging) by $75\times$ from 202.8\,kSa/s to 2.704\,kSa/s. Finally, spurious signals at 49\,Hz, 50\,Hz, and 60\,Hz and remaining cross-talk at 338\,Hz not completely eliminated by the band-pass filter are removed with 1-Hz-wide FFT notch-stop filters. The resulting signals each have single-sided equivalent noise bandwidth $f_\text{ENBW} \!=\! 203$\,Hz~\cite{Shelton1970, Winder2002}.

\subsection{Lock-In Signal Slope Measurement} \label{slope}

The demodulated lock-in signals $S_i$ for channels $i \!=\! \lambda, \chi, \varphi, \kappa$ are converted to frequency shifts using the following calibration: to each channel (sequentially), a 20\,kHz frequency chirp is applied to the modulated MWs. The PL voltage signal is measured, downmixed, and filtered as described in the main text, and the resulting signal is plotted vs.~MW frequency and fit to a line, from which the slope is extracted.  The slope for each channel in $\upmu$V/kHz is averaged over $\sim\!25$~seconds. For the data shown in Fig.~\ref{fig:nvshifts}, the measured slopes for the four channels are tabulated in Table~\ref{tab:params}.

\section{Shot-Noise-Limited Vector Field Sensitivity}\label{shot}
The photon-shot-noise-limited sensitivities to magnetic fields oriented along $\hat{x}$, $\hat{y}$, and $\hat{z}$ are calculated herein~\cite{Jaynes2003,Bendat2010,Goodman2015}. First we consider only photon shot noise on the detected PL signal from the diamond. The smallest detectable ODMR line shift (SNR = 1) due to a magnetic field in the presence of shot noise from photoelectrons on the signal photocurrent alone, $\Delta \nu_{i,\text{min}}^{I_\text{sig}}$, is given by 
\begin{equation}\label{eqn:shot}
\Delta \nu_{i,\text{min}}^{I_\text{sig}} =\sigma_i^{I_\text{sig}} =\frac{R_\text{sig}}{\frac{dS_i}{d\Delta \nu_i}}\sqrt{2 \,q \,I_\text{sig}\Delta f},
\end{equation}
where $\sigma_i^{I_\text{sig}}$ is the standard deviation of the frequency shifts (in Hz) on channel $i$ due to shot noise on the PL photocurrent, $R_\text{sig}$ is the signal photodiode's termination in ohms, $\frac{dS_i}{d\Delta \nu_i}$ is the PL lock-in signal slope in V/Hz, 
$q$ is the elementary charge, and $\Delta f$ is the single-sided measurement bandwidth. 

When limited by photon shot noise, fluctuations on each of the four lock-in detection channels are uncorrelated, and the covariance matrix of ODMR frequency fluctuations is given by    
\begin{equation}\label{eqn:covariance}
\bm{\Sigma}_{\Delta \nu} =
\begin{bmatrix}
(\sigma_\lambda^{I_\text{sig}})^2 & 0 & 0 & 0\\
0 & (\sigma_\chi^{I_\text{sig}})^2 & 0 & 0\\
0 & 0 & (\sigma_\varphi^{I_\text{sig}})^2 & 0\\
0 & 0 & 0 & (\sigma_\kappa^{I_\text{sig}})^2\\
\end{bmatrix}.
\end{equation}
The sensitivities to fields along $\hat{x}$, $\hat{y}$, and $\hat{z}$ are found by transforming $\bm{\Sigma}_{\Delta \nu}$ into a covariance matrix of lab-frame magnetic field shifts:
\begin{equation}\label{eqn:covariancetransform}
\bm{\Sigma}_B = \textbf{A}^+\bm{\Sigma}_{\Delta \nu} \left(\textbf{A}^{+}\right)^\text{T},
\end{equation} 
where $\left(\textbf{A}^{+}\right)^\text{T}$ is the transpose of the matrix $\textbf{A}^{+}$. The diagonal elements of $\bm{\Sigma}_{B}$ are the variances  $(\sigma_x^{I_\text{sig}})^2$, $(\sigma_y^{I_\text{sig}})^2$, and $(\sigma_z^{I_\text{sig}})^2$, which have dimensions of tesla$^2$. Off-diagonal elements of $\bm{\Sigma}_{B}$ represent noise correlations; for diagonal $\bm{\Sigma}_{\Delta \nu}$, these correlations arise from mixing of the $|m_s\!=\!\pm 1\rangle$ states with the $|m_s\!=\!0 \rangle$ state due to the bias magnetic field's non-negligible projections transverse to the NV symmetry axes. 

The sensitivity $\eta_x^{I_\text{sig}}$ is given by
\begin{equation}\label{eqn:etashot}
\eta_x^{I_\text{sig}} = \sigma_x^{I_\text{sig}} \sqrt{T},
\end{equation}
where, for continuous readout as in CW-ODMR, the measurement time is $T \!=\! 1/(2\Delta f)$. Note this definition matches Eqn.~\ref{eqn:sensitivity}, where $\Delta f \!=\! f_\text{ENBW}$, and with $\sigma_x^{I_\text{sig}}$ here replacing the measured standard deviation $\sigma_x$. The y- and z-sensitivities $\eta_y^{I_\text{sig}}$ and $\eta_z^{I_\text{sig}}$ are defined equivalently.

As described in Appendix~\ref{noisecancellation}, the present experiment approaches shot-noise-limited sensing by canceling laser intensity fluctuations using a reference photodetector and a software-based noise cancellation protocol. Here we consider the limit wherein this method completely cancels laser intensity fluctuations. In this limit, shot noise from the reference photocurrent also contributes to the estimated sensitivity. Including this contribution as a separate uncorrelated noise source increases the minimum detectable ODMR line shift by a term $\mathscr{F}_\text{ref} \!=\! \sqrt{1+\frac{I_\text{sig}}{I_\text{ref}}}$ to $\Delta \nu_{i,\text{min}}^{I_\text{sig}, I_\text{ref}} \!=\! \mathscr{F}_\text{ref} \Delta \nu_{i,\text{min}}^{I_\text{sig}}$. In the limit of high reference photocurrent, $\mathscr{F}_\text{ref}$ approaches 1; when $I_\text{sig} \!=\! I_\text{ref}$, $\mathscr{F}_\text{ref}\!=\!\sqrt{2}$. The term $\mathscr{F}_\text{ref}$ enters Eqns.~\ref{eqn:covariance}-\ref{eqn:etashot} in the form of a constant prefactor, such that the photon-noise-limited sensitivity, modified by the reference detection, is given by $\eta_x^{I_\text{sig},I_\text{ref}} \!=\! \mathscr{F}_\text{ref} \sigma_x^{I_\text{sig}}\sqrt{T}$.

In the present experiment, the reference photodiode collects an average photocurrent $I_\text{ref} \!=\! 30.1$\,mA from the picked-off 532\,nm beam, and the signal photodiode collects an average photocurrent $I_\text{sig} \!=\! 24.1$\,mA from the diamond PL. Inserting these values into Eqn.~\ref{eqn:shot} along with $R_\text{sig} \!=\! 300\,\Omega$ and the slopes $\frac{dS_i}{d\Delta \nu_i}$ given in Table~\ref{tab:params} yields, for a 1-second measurement (i.e., $\Delta f \!=\! 0.5$\,Hz): $\Delta \nu_{\lambda,\text{min}}^{I_\text{sig}, I_\text{ref}} \!=\! 0.632$\,Hz,  $\Delta \nu_{\chi,\text{min}}^{I_\text{sig}, I_\text{ref}} \!=\! 0.594$\,Hz, $\Delta \nu_{\varphi,\text{min}}^{I_\text{sig}, I_\text{ref}} \!=\! 0.468$\,Hz, and $\Delta \nu_{\kappa,\text{min}}^{I_\text{sig}, I_\text{ref}} \!=\! 0.600$\,Hz. Using Eqns.~\ref{eqn:covariance}-\ref{eqn:etashot} with $T \!=\! 1$\,s to calculate the photon-noise-limited sensitivities along $\hat{x}$, $\hat{y}$, and $\hat{z}$, we find $\eta_x^{I_\text{sig}, I_\text{ref}} \!=\! 18.1$\,pT/$\sqrt{\text{Hz}}$, $\eta_y^{I_\text{sig}, I_\text{ref}} \!=\! 18.4$\,pT/$\sqrt{\text{Hz}}$, and $\eta_z^{I_\text{sig}, I_\text{ref}} \!=\! 17.5$\,pT/$\sqrt{\text{Hz}}$, which are 2.5\,-\,3$\times$ better than the realized sensitivities of the present device.

\bibliography{SVMbib2017.bib}

\end{document}